\documentclass[aip,jcp,preprint,graphicx]{revtex4-1}
\usepackage{graphicx}% Include figure files
\usepackage{amssymb}
\usepackage[version=3]{mhchem}
\usepackage{siunitx}
\usepackage{hyperref}
\DeclareSIUnit\torr{torr}
\draft % marks overfull lines with a black rule on the right

\begin{document}

%\preprint{}
\newcommand{\gnew}{9775.0018(45)}
\newcommand{\bnew}{1.162222(37)}
\newcommand{\dnew}{3.998(62)}
\newcommand{\Gnew}{\SI{\gnew}{\per\cm}}
\newcommand{\Bnew}{\SI{\bnew}{\per\cm}}
\newcommand{\Dnew}{\SI{\dnew d-6}{\per\cm}}
\newcommand{\GNold}{9663.3385(10)}
\newcommand{\BNold}{1.158021(10)}
\newcommand{\DNold}{1.659(20)}
\newcommand{\Gold}{\SI{\GNold}{\per\cm}}
\newcommand{\Bold}{\SI{\BNold}{\per\cm}}
\newcommand{\Dold}{\SI{\DNold d-6}{\per\cm}}
\newcommand{\newstate}{$\nu_1+\nu_2+\nu_3+\nu_4^1+\nu_5^{-1}$}
\newcommand{\oldstate}{$\nu_1+2\nu_3$}
\newcommand{\sym}{($\Sigma_g^+$)}
\newcommand{\antisym}{($\Sigma_u^+$)}

\title{Double resonant absorption measurement of acetylene symmetric vibrational states probed with cavity ring down spectroscopy} %Title of paper



\author{J. Karhu}
%\email[Corresponding author:]{@helsinki.fi}
%\thanks{}
\affiliation{Laboratory of Physical Chemistry, Department of Chemistry, University of Helsinki, P.O. Box 55, FI-00014 University of Helsinki, Finland}

\author{J. Nauta}
%\thanks{}
%\altaffiliation[Currently at: ]{MPIK Max Planck Institute for Nuclear Physics, Heidelberg, Germany}
\affiliation{Van Swinderen Institute for Particle Physics and Gravity, University of Groningen, Nijenborgh 4, 9747 AG Groningen, Netherlands}
\affiliation{Currently at: MPIK Max-Planck-Institut f{\"u}r Kernphysik, Saupfercheckweg 1, 69117 Heidelberg, Germany}

\author{M. Vainio}
%\thanks{}
\affiliation{Laboratory of Physical Chemistry, Department of Chemistry, University of Helsinki, P.O. Box 55, FI-00014 University of Helsinki, Finland}
\affiliation{VTT Technical Research Centre of Finland Ltd, Centre of Metrology MIKES, P.O. Box 1000, FI-02044 VTT, Finland}
\author{M. Mets{\"a}l{\"a}}
%\thanks{}
\affiliation{Laboratory of Physical Chemistry, Department of Chemistry, University of Helsinki, P.O. Box 55, FI-00014 University of Helsinki, Finland}
\author{S. Hoekstra}
%\email[Corresponding author:]{@helsinki.fi}
%\thanks{}
\affiliation{Van Swinderen Institute for Particle Physics and Gravity, University of Groningen, Nijenborgh 4, 9747 AG Groningen, Netherlands}
\author{L. Halonen}
%\thanks{}
\affiliation{Laboratory of Physical Chemistry, Department of Chemistry, University of Helsinki, P.O. Box 55, FI-00014 University of Helsinki, Finland}
\noindent This article may be downloaded for personal use only. Any other use requires prior permission of the author and AIP Publishing.\\
The following article appeared in:\\
Karhu, J. \emph{et al.}, J. Chem. Phys. \textbf{144}, 244201 (2016)\\
and may be found at \url{http://dx.doi.org/10.1063/1.4954159}
\date{\today}
\begin{abstract}
%A novel mid-infrared/near-infrared double resonant absorption setup for accessing symmetric vibrational state of acetylene is presented.
A novel mid-infrared/near-infrared double resonant absorption setup for studying infrared-inactive vibrational states is presented. A strong vibrational transition in the mid-infrared region is excited using an idler beam from a singly resonant continuous-wave optical parametric oscillator, to populate an intermediate vibrational state. High output power of the optical parametric oscillator and the strength of the mid-infrared transition result in efficient population transfer to the intermediate state, which allows measuring secondary transitions from this state with a high signal-to-noise ratio. A secondary, near-infrared transition from the intermediate state is probed using cavity ring-down spectroscopy, which provides high sensitivity in this wavelength region. Due to the narrow linewidths of the excitation sources, the rovibrational lines of the secondary transition are measured with sub-Doppler resolution. The setup is used to access a previously unreported symmetric vibrational state of acetylene, \newstate~in the normal mode notation. Single-photon transitions to this state from the vibrational ground state are forbidden. Ten lines of the newly measured state are observed and fitted with the linear least-squares method to extract the band parameters. The vibrational term value was measured to be at \Gnew, the rotational parameter $B$ was \Bnew, and the quartic centrifugal distortion parameter $D$ was \Dnew, where the numbers in the parenthesis are one-standard errors in the least significant digits.

\end{abstract}

\pacs{}% insert suggested PACS numbers in braces on next line

\maketitle %\maketitle must follow title, authors, abstract and \pacs

\section{Introduction}
%\label{}
For a linear molecule with an inversion center, transitions from the vibrational ground state to other symmetric vibrational states are infrared-inactive within the ground electronic state. A well-studied example of such a molecule is acetylene (\ce{C2H2}). The transitions to the symmetric states of acetylene can be measured with Raman spectroscopy \cite{RefWorks:229}, as hot bands\cite{RefWorks:230} or with multiphoton methods.\cite{RefWorks:231} In the first two cases, the transitions to high-energy vibrational overtones and combination states are generally too weak to be measured. The standard way to measure the high-energy symmetric vibrational states is using two photon methods, where molecules are first brought to the lowest excited electronic state and laser induced fluorescence (LIF) to a vibrational state within the ground electronic state is measured.\cite{RefWorks:232} Methods based on LIF alone are often limited in resolution by the equipment needed to disperse the fluorescence and ultimately by the Doppler-profile of the transition. Stimulated emission pumping (SEP), where changes in the fluorescence are observed while a laser depopulates the intermediate state by stimulated emission, relaxes these limits.\cite{RefWorks:233} However, transitions from the excited electronic state can limit the detectable states to those where a high number of trans-bend and CC stretch quanta are excited, due to the geometry of the excited electronic state.\cite{RefWorks:233}

In our previous work, we have used LIF setups to measure high-energy symmetric stretching states while remaining in the electronic ground state.\cite{RefWorks:244,RefWorks:239} The acetylene sample was optically pumped to a high-energy anti-symmetric vibrational state inside a laser cavity and the resulting LIF was dispersed using a high-resolution Fourier-transform infrared spectrometer. 
Transitions to high-energy vibrational states are much weaker than electronic transitions, resulting in a relatively low signal-to-noise ratio (SNR). In a later work, with careful optimization of the fluorescence gathering optics, the SNR was improved significantly.\cite{RefWorks:238} In a more recent study, we developed a stimulated emission scheme, where the intermediate state was a high-energy vibrational state on the electronic ground state. The setup could reach a superior SNR compared to the LIF measurements, but it relied on direct observation of the stimulated emission and required a rigorous locking scheme to reach a large enough population on the intermediate state.\cite{RefWorks:234} 

\begin{figure}
\includegraphics{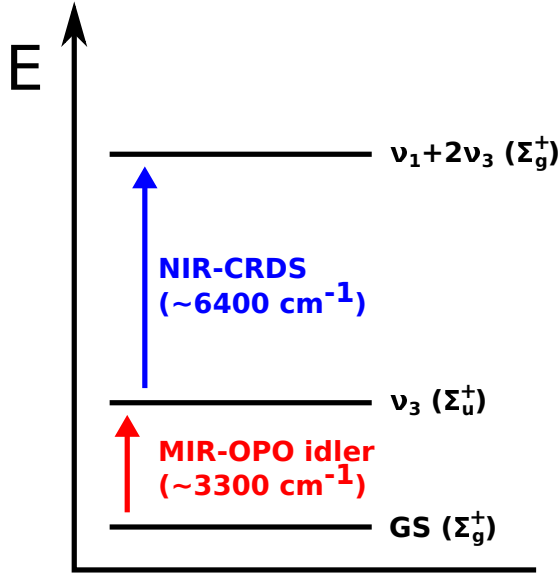}%[width=8.5cm]
\caption{\label{ediag}Energy diagram for one of the measured transitions in acetylene (\ce{C2H2}). An idler beam from a mid-infrared optical parametric oscillator (MIR-OPO) excites the molecule to the fundamental excited state of the antisymmetric CH stretching vibration ($\nu_3$). The second, weaker transition is probed using near-infrared cavity ring down spectroscopy (NIR-CRDS). The symbol $\nu_1$ denotes the symmetric CH stretching vibration and the symbol GS stands for the vibrational ground state.}%
\end{figure}

Here, we present a novel way of measuring symmetric vibrational states, based on incoherent two-photon absorption. We first excite the molecules to an anti-symmetric intermediate state using a mid-infrared beam from an optical parametric oscillator (OPO). Then, we probe a second transition to the final symmetric state using near-infrared cavity ring down spectroscopy (CRDS) (figure \ref{ediag}). The mid-infrared region contains a very strong fundamental excitation of the anti-symmetric stretching vibration, which, together with the high output power of the OPO, results in efficient optical pumping and a high population transfer to the intermediate state. The second transition lies in the near-infrared region, where CRDS performs well, due to highly reflective cavity mirrors and well developed detectors available in this wavelength region. With this simple and robust measurement system, we can reach an SNR that surpasses our previous dispersed LIF experiments\cite{RefWorks:238} and is comparable to our stimulated emission measurement, which used a significantly more complicated laser setup.\cite{RefWorks:234} Like the stimulated emission measurement, the current setup also provides sub-Doppler resolution.

Acetylene is a molecule of substantial interest, not only in high-resolution spectroscopy, but also in computational chemistry and astrochemistry.\cite{RefWorks:228,RefWorks:227,RefWorks:226} It has been detected in many astronomical environments and is thought to play an important role in many processes of the carbon astrochemistry, such as formation of polycyclic aromatic hydrocarbons and the growth of carbon chains.\cite{RefWorks:253,RefWorks:254} In laboratory spectroscopy, it is often used as a model species, since its simple, linear geometry gives its rovibrational spectra a regular and assignable structure. This makes it feasible to study the large number of couplings and resonances that affect the spectra. The infrared spectrum of acetylene can be modeled to a high precision using a cluster model.\cite{RefWorks:243} The cluster model is based on fitting the model parameters on experimental data. It relies on availability of high quality experimental values for the state energies to improve the modeling of the various coupling mechanisms. The states with mainly stretching contribution can also be understood with the considerably more simple local-mode picture.\cite{RefWorks:246} In this article we have used our new double-resonance method to access a previously unobserved symmetrical vibrational state of acetylene.

\section{Experimental Technique}
\begin{figure*}
\includegraphics{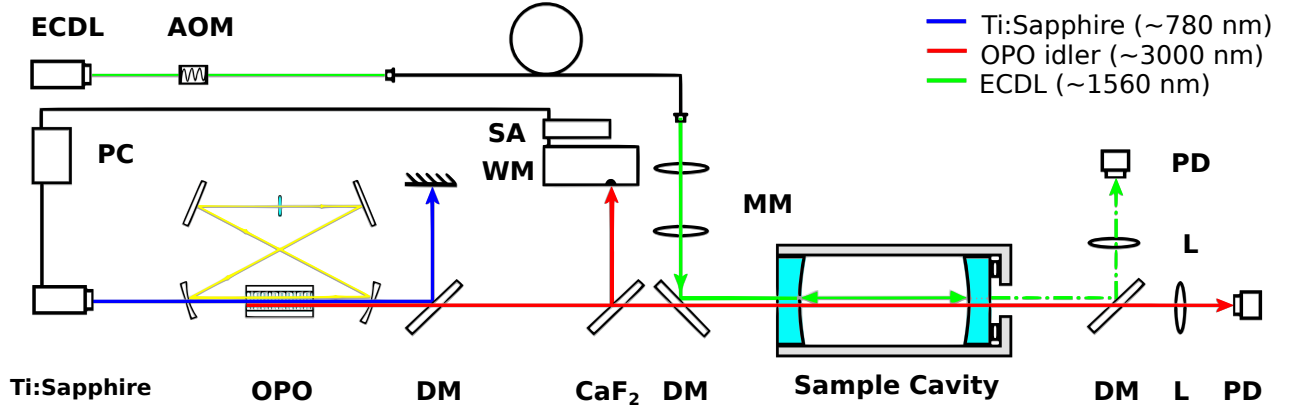}%[width=17cm]
\caption{\label{scheme}Schematic overview of the measurement setup. A titanium-sapphire ring laser (Ti:Sapphire) is used to pump an optical parametric oscillator (OPO). The mid-infrared idler beam of the OPO is separated from the residual pump and signal beams using a dichroic mirror (DM). The idler beam passes through the sample cavity and is focused on a photo detector (PD) with an anti-reflection coated lens (L). A small portion of the idler power is reflected to a wavelength meter (WM) using a calcium fluoride window (\ce{CaF2}). The WM signal is passed through a spectrum analyser (SA) to a computer (PC), where a slow controller tunes the OPO pump wavelength to stabilize the wavelength of the mid-infrared idler beam. An external cavity diode laser (ECDL) is directed to an acousto-optical modulator (AOM). The first diffraction order from the AOM is brought to the measurement system using an optical fiber. The fiber output is mode matched (MM) to the sample cavity. The sample cavity is a low-expansion cavity connected to a gas line, which can be used to control the acetylene sample pressure at millitorr level. The rear mirror of the sample cavity is attached to a piezo actuator. The light leaking out of the cavity through the rear mirror is focused on a photo detector. A small portion of the ECDL power is split just after the laser head for wavelength measurement (not shown).}%
\end{figure*}

The schematic overview of the setup is presented in figure \ref{scheme}. The mid-infrared transition is excited using an idler beam from an in-house built continuous wave (CW) optical parametric oscillator.\cite{RefWorks:224} The OPO is pumped using a titanium-sapphire ring laser (MBR-PS, Coherent), which is in turn optically pumped at \SI{532}{\nm} by a high power CW-laser (Verdi V-18, Coherent). The OPO cavity has four mirrors in a bow-tie configuration. The mirrors are highly reflective at the signal wavelength of about \SI{1}{\micro\m}. The idler beam, with a wavelength tunable between \num{2.5} and \SI{4.4}{\micro\m}, has an output power of up to \SI{100}{\mW} around the wavelength of \SI{3}{\micro\m}, which was used in the measurements. The non-linear medium is a \SI{5}{\cm} long periodically poled lithium niobate crystal, doped with magnesium oxide (\ce{MgO}:PPLN, HC Photonics). The OPO cavity includes an intracavity YAG etalon to ensure single mode operation and to reduce mode hops. After the cavity, a small portion of the idler output power is reflected for a wavelength measurement using a \ce{CaF2}-plate. The wavelength is monitored using a wavemeter (WA-1500-IR, EXFO) connected to a spectrum analyzer (WA-650, EXFO). The transmitted beam is directed to the sample cavity which it traverses only once, exciting a strong transition to an anti-symmetric state of acetylene at about \SI{3300}{\per\cm}. After the cavity, a photodetector (PVI-2TE-5, VIGO) measures the transmitted idler power. The sample cavity is \SI{35}{\cm} long and made out of glass with a low thermal expansion coefficient. The cavity is connected to gas line system with an acetylene supply, a vacuum oil pump (Edwards) and a low pressure capacitance manometer (Baratron, MKS Instruments), which are used to control the pressure of the acetylene sample within the cavity.

The transition from the anti-symmetric state to the final symmetric state is probed using CW-CRDS.\cite{RefWorks:225} The laser source is an external cavity diode laser (ECDL, New Focus, Velocity 6328) with a wavelength tunable between \num{1520} and \SI{1570}{\nm}. The ECDL provides up to \SI{5}{\mW} of power, as measured just before the beam enters the sample cavity. The wavelength of the ECDL is measured continuously with a wavemeter (WA-1500-NIR, EXFO). The ECDL beam is sent through an acousto-optic modulator, which is used to initiate the ring down events, and the first order diffracted beam is mode matched to the TEM$_{00}$ –mode of the sample cavity. The transmitted power after the cavity is measured with a photodetector (D100, RedWave Labs) and the detector signal is passed to a LabView (National Instruments) program for signal processing and extraction of the ring down times. The cavity losses are determined as the reciprocal of the ring down times, divided by the speed of light. The rear mirror of the cavity is fixed to a piezo actuator, which is periodically scanned over one free spectral range of the cavity with a frequency of \SI{15}{\Hz}. The cavity mirrors possess reflectivity of about \num{0.99998} at the ECDL wavelength, resulting in empty cavity ring down times of about \SI{60}{\micro\s}.

The transition lines are measured one by one. The idler frequency is tuned on top of one of the rovibrational lines in the mid-infrared region. To counteract long term drifts, the idler wavelength is kept fixed using a slow proportional-integral (PI) feedback loop running in a LabView-environment. The idler wavelength is controlled by tuning the MBR-PS wavelength. The time constant of the feedback loop is in the order of a few seconds. While the idler wavelength is fixed on the mid-infrared transition, the ECDL wavelength is scanned over the near-infrared one. The ECDL is scanned in steps of about \SI{15}{\MHz}. The step size is limited by the resolution of the wavemeter and the long term linewidth of the ECDL. At each step, 50 ring down signals are averaged, which takes up to \SI{2}{\s}. The background was determined by scanning over the transition while the idler beam was blocked from entering the sample cavity. The absorption is determined as the difference in cavity losses between a sample scan and a background scan.

\section{Results and discussion}
The experimental setup described was used to measure two symmetric states. The state \oldstate~\sym, where $\nu_1$ and $\nu_3$ represent the symmetric and anti-symmetric CH stretching vibrations, respectively, was used for development and testing of the measurement system. For this state, the idler beam first excites the mid-infrared transition to the fundamental state of the anti-symmetric stretch $\nu_3$. The CRDS then probes the transition \oldstate~$\leftarrow$ $\nu_3$. The setup was also used to observe the state \newstate~\sym, where $\nu_2$, $\nu_4$ and $\nu_5$ represent the CC stretching, symmetric bending and antisymmetric bending vibrations, respectively. The superscripts denote the value of the vibrational angular momentum. To our knowledge, this state has not been accessed before.  For this state, the idler beam excites the transition to the state $\nu_2+\nu_4^1+\nu_5^{-1}$ \antisym. The otherwise weak transition to this combination state is enhanced by a strong anharmonic vibrational resonance with the $\nu_3$-fundamental state, such that the transitions to these two states are of a similar intensity.\cite{RefWorks:242} The transition \newstate~$\leftarrow$ $\nu_2+\nu_4^1+\nu_5^{-1}$ is then probed using CRDS. For both bands measured here, the setup reaches a signal-to-noise ratio, defined here as the peak height divided by the standard deviation of the background signal in the CRDS spectrum, of about 300. The state \oldstate~has been measured before using dispersed LIF\cite{RefWorks:238} and infrared stimulated emission probing (IRSEP).\cite{RefWorks:234} For this state, our setup provides a better SNR than the LIF measurement, where the SNR was about 100 at best, and is comparable to the significantly more complicated IRSEP system, which reached an SNR of over 500.

\begin{figure}
\includegraphics{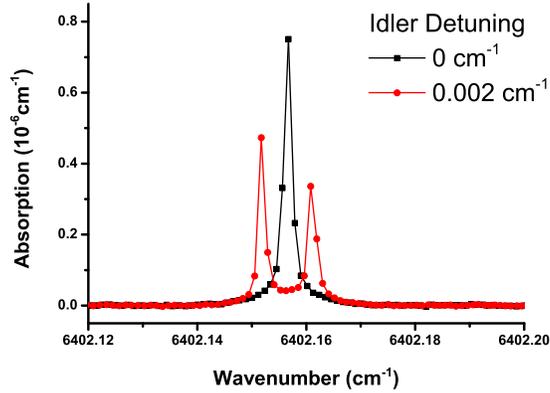}%
\caption{\label{offset}Two overlaid CRDS spectra of the near-infrared transition (line R(15) of the transition \oldstate~$\leftarrow$ $\nu_3$) with different idler wavelengths. The highest peak at the center (black/$\blacksquare$) corresponds to the idler wavelength being at the center of the Doppler-profile of the mid-infrared transition (line R(15) of the transition $\nu_3$ $\leftarrow$ GS). For the other spectrum (red/$\bullet$), the idler wavelength has a small offset from the center value.}%
\end{figure}

The transition from one of the anti-symmetric states of the resonance dyad ($\nu_3$/$\nu_2+\nu_4^1+\nu_5^{-1}$) to the final symmetric state can be observed as a sub-Doppler line in the CRDS spectrum. The sub-Doppler profile is the result of the idler beam only exciting molecules for which the Doppler shifted transition frequency lies within the narrow idler linewidth. The speed distribution of the molecules evolves little while on the intermediate state, so the ECDL probes molecules with that same initial Doppler-shift. If the idler wavelength is detuned from the center wavelength of the first transition, the second transition appears as two separate sub-Doppler peaks, located symmetrically around the center wavelength of the Doppler broadened profile (figure \ref{offset}). This is due to the ECDL forming two counterpropagating beams within the linear cavity, so that its wavelength matches the Doppler-shift of the excited molecules on both sides of the Doppler profile, when scanned over it. This can be used to accurately find the center wavelength of the mid-infrared transition, as only when the idler wavelength matches the molecules with near-zero Doppler shift, the near-infrared transition appears as a single peak in the CRDS spectrum. 

\begin{figure}
\includegraphics{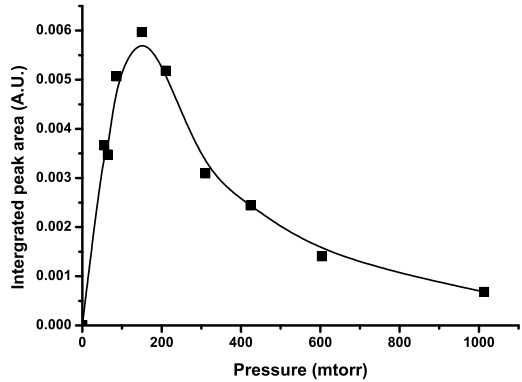}%
\caption{\label{pdep}Signal intensity, calculated as the integrated peak area, of the cavity ring down spectrum of the line R(16) of the transition \oldstate~$\leftarrow$ $\nu_3$, when measuring the same line at different acetylene pressures. The measurement points ($\blacksquare$) have been fitted with basic splines (\textbf{--}).}%
\end{figure}

The optimal sample pressure was determined by measuring the CRDS signal strength as a function of acetylene pressure, while keeping the OPO power constant (figure \ref{pdep}). The signal strength increases first as the concentration of acetylene increases, but starts to drop after a certain pressure value, as the life time of the intermediate state starts to decrease due to a larger number of collisions. The optimal pressure was found to lie between \num{100} and \SI{200}{\milli\torr}.

\begin{figure}
\includegraphics{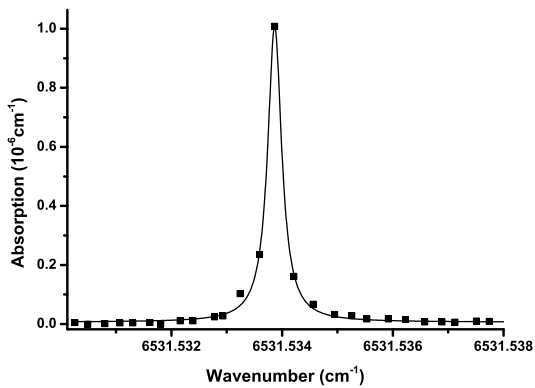}%
\caption{\label{lorentz}A measured cavity ring down spectrum for the line R(17) of the transition between the intermediate state $\nu_2+\nu_4^1+\nu_5^{-1}$ and the final state \newstate. The measurement points ($\blacksquare$) have been fitted with a Lorentzian line profile (\textbf{--}).}%
\end{figure}

The sub-Doppler peaks were fitted with the Lorentzian line profile to find the center wavelength of the transition (figure \ref{lorentz}). The linewidths were typically about \SI{10}{\MHz} (the full width at half maximum). The expected pressure self-broadened linewidths are only a few \si{\MHz} in this pressure range\citep{RefWorks:235}, so the measured linewidths are likely limited by the idler linewidth and wavelength stability. Because the ECDL linewidth is less than \SI{5}{\MHz} on the time scale of seconds, it is not the limiting factor, although it may also affect the measured transition linewidth. The lines do not always fit well to the Lorentzian profile, which is likely partially because the linewidths are already near the resolution limits of our wavemeter and partially because of fluctuations in the idler power and wavelength during the scan over the near-infrared transition. The speed distribution of the molecules may also evolve slightly during the lifetime of the intermediate state.

By measuring the spectrum of the same near-infrared transition with varying idler wavelengths, the line center of the mid-infrared transition can be determined by the appearance of only a single peak on the CRDS spectrum (see figure \ref{offset}). The idler wavelength can then be compared to well-known line positions of the resonance-dyad.\cite{RefWorks:242} The wavemeter used to measure the ECDL wavelength was calibrated using known line positions of the $\nu_1+\nu_3$ transition of acetylene, which have been measured accurately by NIST.\cite{RefWorks:236} Known atmospheric water lines\citep{RefWorks:237} were used for the calibration as well, to cover the whole used spectral range and to avoid problems with extrapolating the calibration outside of the $\nu_1+\nu_3$ band. The calibration was checked regularly during the several months of development and measurements. On average, the offset of the wavemeter was \SI{0.0046}{\per\cm}, which was used to correct the line positions. For the worst case, the deviation from the average was \SI{0.0061}{\per\cm}, which can then be considered as the upper limit for the accuracy of the measured line positions. For the actual measurements of the sub-Doppler lines of the state \oldstate, multiple measurements of the same line at different points in time during the development process resulted in a standard deviation of \SI{0.0027}{\per\cm} in the line position, for the worst case.

For the state \oldstate, 9 lines measured with the optimized setup were included in the final fit. The energies of the rotational states within the final symmetric vibrational state were calculated as a sum of the rotational energy of the vibrational ground state\cite{RefWorks:241}, a reference value for the transition to the resonance dyad\cite{RefWorks:242} and the measured transition energy between the resonance dyad and the final symmetric state. To extract the rotational parameters, the rovibrational energies were fitted using the linear least squares method. The rovibrational energy was expressed as:
\begin{equation}
E/hc=G+BJ(J+1)-DJ^2(J+1)^2+...
\end{equation}
where $E$ is the measured energy, $G$ is the vibrational term value, $B$ is the rotational parameter, and $D$ is the quartic centrifugal distortion parameter. For the ground state, the sextic centrifugal distortion parameter $H$ was also included, but for all other states, only terms up to the quartic centrifugal distortion were found significant. The vibrational term value derived from the fit was \Gold, where the value in parenthesis is one-standard error in the least significant digits. Previous measurements of this state give values of \SI{9663.3860(11)}{\per\cm} (ref. \citenum{RefWorks:238}), \SI{9663.3508(57)}{\per\cm} (ref. \citenum{RefWorks:239}) and \SI{9663.362(16)}{\per\cm} (ref. \citenum{RefWorks:234}) for the vibrational term value. The other parameters from the fit are presented in table \ref{t_oldband}. Although the statistical uncertainty is the smallest for the first reference (ref. \citenum{RefWorks:238}), it differs somewhat from our result. The other two (refs \citenum{RefWorks:239} and \citenum{RefWorks:234}) agree well with our result, falling within three standard deviations from our value. The difference to the first reference may be due to a calibration offset in addition to the statistical errors, as the parameters $B$ and $D$ agree well within three standard deviations. However, our calibration measurements suggest that the inaccuracy of our line measurements is not more than \SI{0.007}{\per\cm} and therefore would not by itself explain the difference. There may be some irregularities in the individual line positions, caused by $J$-dependent perturbation effects, which could lead to a difference in the band parameters when measuring a larger number of lines, as was the case in the first reference.

\begin{table*}
\caption{\label{t_oldband}The rotational parameters for the vibrational state \oldstate~.}
\begin{ruledtabular}
\begin{tabular}{c|cccc}
Param.\footnotemark[1] &This work\footnotemark[2]&Ref.\citenum{RefWorks:238}&Ref.\citenum{RefWorks:239}&Ref.\citenum{RefWorks:234}\\
\hline
$G$ (\si{\per\cm}) &\GNold & 9663.3860(11) & 9663.3508(57) & 9663.362(16)\\
$B$ (\si{\per\cm}) &\BNold & 1.158053(16) & 1.15814(14) &---\footnotemark[3]\\
$D$ (\SI{d-6}{\per\cm}) &\DNold & 1.742(38) & 1.54(72) &---\footnotemark[3]\\
\hline
$N$\footnotemark[4] & 9 & 35 & 14  & 8
\end{tabular}
\footnotetext[1]{$G$ is the vibrational term value, $B$ is the rotational parameter and $D$ is the quartic centrifugal distortion parameter.}
\footnotetext[2]{Values in parenthesis are one-standard error in the least significant digits.}
\footnotetext[3]{Not reported}
\footnotetext[4]{Number of spectral features used in the least-squares fit}
\end{ruledtabular}
\end{table*}

The rovibrational energies for the state \newstate~were calculated as described above for the state \oldstate. Ten lines of the new state were measured (table \ref{lines}) and the energies were fitted with the linear least squares method (table \ref{t_newband}). The vibrational term value was \Gnew, $B$ was \Bnew and $D$ was \Dnew. To our knowledge, this state has not been observed previously, but the results can be compared to values calculated using the cluster model, based on a global fit of observed vibrational states.\cite{RefWorks:243} The model predicts the values \SI{9774.2}{\per\cm}, for the vibrational term value, and \SI{1.162}{\per\cm}, for the rotational parameter.\cite{RefWorks:240} The values agree well with our results, as the vibrational terms differ by less than \SI{1}{\per\cm}. The agreement with the cluster model makes us confident in the assignment of the newly measured state. There are no other predicted states, which we would expect to be detectable with our measurement system, with similar values of the vibrational term value and rotational parameter.
\begin{table}
\caption{\label{lines}Measured rovibrational energies for the vibrational state \newstate.}
\begin{ruledtabular}
\begin{tabular}{ccc}
%\hline
%\hline
J\footnotemark[1] &$E_{obs.}$ (\si{\per\cm})\footnotemark[2]&$E_{obs.}-E_{calc.}$ (\si{\per\cm})\footnotemark[3]\\
\hline
7&9840.06967&\num{-0.00397}\\
9&9879.56511&\num{-0.00421}\\
10&9902.80429&\num{0.00654}\\
11&9928.34671&\num{0.00138}\\
13&9986.39832&\num{0.0047}\\
15&10053.70258&\num{-0.00201}\\
17&10130.26952&\num{0.0024}\\
18&10172.01108&\num{-0.00274}\\
19&10216.06457&\num{-0.00396}\\
23&10415.33157&\num{0.00188}\\
%\hline
%\hline
\end{tabular}
\end{ruledtabular}
\footnotetext[1]{Rotational quantum number of the rovibrational state}
\footnotetext[2]{Measured rovibrational energies}
\footnotetext[3]{The difference between the measured energies and the linear least squares fit.}
\end{table}
 
\begin{table}
\caption{\label{t_newband}The rotational parameters for the vibrational state \newstate.}
\begin{ruledtabular}
\begin{tabular}{c|cc}
Param.\footnotemark[1] &This work\footnotemark[2]&Calculated\cite{RefWorks:240}\\
\hline
$G$ (\si{\per\cm}) &\gnew & 9774.2\footnotemark[3] \\
$B$ (\si{\per\cm}) &\bnew & 1.162\\
$D$ (\SI{d-6}{\per\cm}) &\dnew &--- 
\end{tabular}
\footnotetext[1]{$G$ is the vibrational term value, $B$ is the rotational parameter and $D$ is the quartic centrifugal distortion parameter.}
\footnotetext[2]{Values in parenthesis are one-standard error in the least significant digits.}
\footnotetext[3]{At around these energies, the vibrational term values calculated with the cluster model differ from experimental values by up to about \SI{1}{\per\cm}.}
\end{ruledtabular}
\end{table}

Accessing symmetric vibrational states provides results which can be compared to a simple local-mode model, which relates the CH stretching states of different symmetry.\cite{RefWorks:246} The state \oldstate~is [$30^+$] in local-mode picture, where the two numbers within the brackets refer to the number of vibrational excitations in each CH bond and the sign refers to the symmetry of the state. There is a corresponding anti-symmetric state [$30^-$], or $3\nu_3$ in the normal mode picture. According to the local-mode theory, if the two CH bond oscillators were uncoupled, the vibrational term values of the local-mode pair would be the same.\cite{RefWorks:246} For states such as these, where only one of the CH bond oscillators is excited, the CH bonds become more uncoupled at higher vibrational excitations, due to the increasing anharmonicity becoming dominant over the coupling between the CH bonds, and the vibrational term values of the local-mode pair approach each other. The same is true for their rotational parameters.\cite{RefWorks:251} The vibrational term value and rotational parameter for the state [$30^-$] are \num{9639.854} and \SI{1.158356}{\per\cm} respectively,\cite{RefWorks:247} so the splitting between the local-mode pair [$30^+$]/[$30^-$] is \SI{23.48}{\per\cm} for the vibrational term values and \SI{0.00034}{\per\cm} for the rotational parameters. For comparison, for a lower energy pair [$20^+$]/[$20^-$] ($2\nu_3$/$\nu_1+\nu_3$ in the normal mode picture), the splittings are indeed larger, with the values \num{42.45} and \SI{0.00175}{\per\cm} for the vibrational term values and the rotational parameters, respectively.\cite{RefWorks:245} The local-mode representation for the newly measured state \newstate~is [$20^+$]1+($\nu_4^1+\nu_5^{-1}$), where the number after the square brackets refers to the number of excitations in the CC bond oscillator. Its counterpart state [$20^-$]1+($\nu_4^1+\nu_5^{-1}$), or $\nu_2+2\nu_3+\nu_4^1+\nu_5^{-1}$ in normal mode picture, has a vibrational term value of \SI{9744.5467(8)}{\per\cm} and a rotational parameter \SI{1.16433}{\per\cm}.\cite{RefWorks:247} The splitting between the vibrational term values is then \SI{30.46}{\per\cm}, and between the rotational parameters it is \SI{0.00211}{\per\cm}. For comparison, the vibrational term value splitting is \SI{29.99}{\per\cm} between the states [$20^+$]+($\nu_4^1+\nu_5^{-1}$) and [$20^-$]+($\nu_4^1+\nu_5^{-1}$),\cite{RefWorks:250,RefWorks:249} so it appears that, in this energy range, addition of one quantum of the CC stretching vibration has little effect on the local-mode behaviour. This fits with the theory, since the indirect coupling of the two CH bonds through the CC bond oscillator should be largely independent of the number of excitations in the CC bond oscillator.\cite{RefWorks:252} The same can be observed when looking at the splitting between the states [$20^+$]1 and [$20^-$]1, which is \SI{44.79}{\per\cm},\cite{RefWorks:239} and the splitting between the states [$20^+$] and [$20^-$], given above as \SI{42.45}{\per\cm}. However, at least in this energy range, the bending vibrations seem to have a more considerable effect on the local-mode behaviour, even though the coupling between the stretching and bending modes is often considered to be weak. One excitation in both bending modes changes the vibrational term value splitting by about \SI{10}{\per\cm}, which is about half of the difference between the splittings of the local-mode pair [$20^+$]/[$20^-$] and the higher energy pair [$30^+$]/[$30^-$].

\section{Conclusions}
We have demonstrated a novel double resonant absorption setup, taking advantage of a strong mid-infrared absorption and the sensitivity of cavity ring down spectroscopy in the near-infrared. While the setup already provides sub-Doppler resolution, the precision of the measurement of the line position might be improved by using a wavelength measurement scheme with better resolution, as we are already working near the limits of the wavemeter we use to measure the ECDL wavelength. This would also allow resolving the line shape of the transition better. However, while the peak profiles could be measured more accurately, the linewidth would still likely be limited by the idler beam's linewidth and long term stability. The pressure broadening limits could possibly be reached by implementing a frequency locking scheme for the OPO to reduce the idler linewidth.

The setup was used to measure two symmetric vibrational states of acetylene near \SI{9700}{\per\cm}. One of the measured states has not been previously accessed. Both transitions were observed with a high SNR and a sub-Doppler resolution. Spectroscopic parameters of the newly observed state were obtained from a linear least squares fit of the measured rovibrational energies, and the results were in good agreement with predicted values from calculations.

Our method is expected to perform well in measuring other high energy symmetric vibrational states of acetylene and its isotopologues. However, the detectable states will likely be limited to those with a low number of excited bending quanta, due to the dominant stretching nature of the states we have used as the intermediate step. In this sense, it can be considered to be complimentary to the methods based on LIF from the excited electronic state. The method should be straightforward to implement for two-photon transitions in other molecules as well, provided that they have a strong infrared-active transition in the mid-infrared region, within the tunable wavelength region of the mid-infrared light source. In our system, the OPO idler wavelength can be tuned between \num{2500} and \SI{4400}{\nm}, which covers, for example, the fundamental transitions of usual CH stretching vibrations. Our ECDL is tunable between \num{1520} and \SI{1570}{\nm}, so, for our system, the second transition needs to lie within this region. In principle, however, the system can be expanded to measure secondary transitions in any wavelength region where CRDS performs well. At longer wavelengths, the limit is usually the availability of high quality optics and light sources. For shorter wavelengths, a higher transition energy generally means a weaker transition strength, so more sensitivity is required from the CRDS system. However, extremely sensitive CRDS setups are constantly developed throughout the infrared region.\cite{RefWorks:255,RefWorks:256}

%\subsection{}
%\subsubsection{}

% If in two-column mode, this environment will change to single-column format so that long equations can be displayed. 
%\begin{widetext}
%$$\mbox{put long equation here}$$
%\end{widetext}

% Figures should be put into the text as floats. 
% Use the graphics or graphicx packages (distributed with LaTeX2e).
% Put the label that you will use with \ref{} command in the braces of the \label{} command.
%
% \begin{figure}
% \end{figure}

% \begin{table}
% \end{tabular}
% \end{table}

\begin{acknowledgments}
We acknowledge the Academy of Finland and the Finnish Funding Agency for Technology and Innovation (TEKES) for funding. J. Karhu is grateful for postgraduate funding from the CHEMS doctoral programme of the University of Helsinki.
\end{acknowledgments}

\section{References}
\bibliography{hcchsymstate}

\end{document}